\documentclass[useAMS,usenatbib]{mn2e}

\usepackage[psamsfonts]{amssymb}
\usepackage[dvips]{graphicx}
\usepackage{amsmath,alltt}
\usepackage{multirow}
\usepackage{rotating}
\usepackage{lscape}

\title[The Evolution of Kicked Black Holes]{The Evolution of Kicked Stellar-Mass Black Holes in Star Cluster Environments}
\author[Webb et al]{Jeremy J. Webb$^{1}$, Nathan W. C. Leigh$^{2}$, Abhishek Singh$^1$, K. E. Saavik Ford$^{2,3,4,5}$, 
\newauthor Barry McKernan$^{2,3,4,5}$, Jillian Bellovary$^{2}$
\thanks{E-mail: jerjwebb@iu.edu (JW); nleigh@amnh.org (NL); sford@amnh.org (KESF); bmckernan@amnh.org (BM); jbellovary@amnh.org (JB)}\\
$^{1}$Department of Astronomy, Indiana University, Swain West, 727 E. 3rd Street, IN 47405 Bloomington, USA \\
$^{2}$Department of Astrophysics, American Museum of Natural History, Central Park West and 79th Street, New York, NY 10024 \\
$^{3}$Department of Science, BMCC, City University of New York, New York, NY 10007, USA\\
$^{4}$Graduate Center, City University of New York, 365 5th Avenue, New York, NY 10016, USA\\
$^{5}$Kavli Institute for Theoretical Physics, UC Santa Barbara, CA 93106, US
}

\begin{document}

\pagerange{\pageref{firstpage}--\pageref{lastpage}} \pubyear{2016}

\maketitle

\label{firstpage}

\begin{abstract}

We consider how dynamical friction acts on black holes that receive a velocity kick while located at the center of a gravitational potential, analogous to a star cluster, due to either a natal kick or the anisotropic emission of gravitational waves during a black hole-black hole merger. Our investigation specifically focuses on how well various Chandrasekhar-based dynamical friction models can predict the orbital decay of kicked black holes with $m_{bh} \lesssim 100 M_\odot$ due to an inhomogeneous background stellar field. In general, the orbital evolution of a kicked black hole follows that of a damped oscillator where two-body encounters and dynamical friction serve as sources of damping. However, we find models for approximating the effects of dynamical friction do not accurately predict the amount of energy lost by the black hole if the initial kick velocity $v_{k}$ is greater than the stellar velocity dispersion $\sigma$. For all kick velocities, we also find that two-body encounters with nearby stars can cause the energy evolution of a kicked BH to stray significantly from standard dynamical friction theory as encounters can sometimes lead to an energy gain. For larger kick velocities, we find the orbital decay of a black hole departs from classical theory completely as the black hole's orbital amplitude decays linearly with time as opposed to exponentially. Therefore, we have developed a linear decay formalism which scales linearly with black hole mass and $\frac{v_{k}}{\sigma}$ in order to account for the variations in the local gravitational potential.

\end{abstract}


\section{Introduction} \label{intro}

Black holes (BHs) are believed to receive velocity kicks both upon formation and as part of a BH-BH merger. When BHs form they receive a natal kick between 0 and 100 km/s, similar to the kicks received by neutron stars \citep{repetto12}. When two black holes (BHs) merge, they experience a kick due to the anisotropic emission of gravitational waves (GWs) \citep[e.g.][]{favata04,merritt04,blecha11}.  The magnitude of the kick can vary by orders of magnitude depending on the binary mass ratio, BH spins and the relative angles of inclination between the BH spin axes and the binary orbital plane.  The kick velocities range from $\lesssim$ 1 km/s to over 500 km/s, reaching a maximum at a mass ratio of q $\sim$ 0.3 \citep{favata04} for maximally misaligned spins.  For very low mass ratios q $\lesssim$ 0.05, the kick velocities are always small $\lesssim$ 100 km/s, independent of the BH spins or their orientations relative to the binary orbital plane (see Figure 1 in \citet{merritt04}).

The four gravitational wave detections to date have been attributed to the merger of BHs between 8 and 36  $M_\odot$ \citep{abbott16, abbott16b, abbott17, abbott17b} which yield remnants with $m_{bh} \lesssim 100 M_\odot$. The dense environments of globular clusters (GCs) represent the most likely host for such events through the merger of dynamically formed BH binaries. Mergers can occur throughout a GC's lifetimes, from their infancy \citep[e.g.][]{leigh13a, leigh13b} to the present-day age of the host galaxy \citep[e.g.][]{rodriguez15, rodriguez16, leigh16}. Nuclear star clusters (NSCs) \citep[e.g.][]{miller08, antonini16} and the massive gas disks in active galactic nuclei (AGN) \citep{mckernan17} have also been suggested as likely locations for stellar BH mergers.  Within these dense environments, BHs should undergo frequent mergers with other BHs \citep{sippel13}. Although the exact details of how intermediate massive BHs (IMBHs) \citep[e.g.][]{bahcall75, portegieszwart04, gratton12, leigh13a, leigh13b,leigh14} and supermassive BH (SMBHs) \citep[e.g.][]{oh02, bromm03, volonteri05, shapiro05, wise08, shang10, tanaka14, madau01,haiman01,volonteri06,tanaka09} form and grow are poorly understood. 

An abundance of observational evidence now exists in favour of the actual existence of BHs in these dense stellar environments. However, the observational evidence in favour of stellar-mass BHs and SMBHs is more compelling than for IMBHs. For example, in the giant elliptical galaxy NGC 4472 in the Virgo Cluster, \citet{maccarone07} reported an accreting BH in an associated GC.  The X-ray luminosity is so high that the authors argue it cannot be anything other than a BH. Shortly thereafter, \citet{shih10} reported an accreting BH in a GC hosted by the giant elliptical galaxy NGC 1399 at the centre of the Fornax Cluster.  As for stellar-mass BHs in GCs, \citet{strader12} recently reported the  detection of two flat-spectrum radio sources in the Galactic GC M22.  If confirmed, these two detections could imply the presence of more unseen BHs, somewhere in the range $\sim 5 - 100$ in M22.  Even more recently, \citet{chomiuk13} reported a candidate BH X-ray binary in the Galactic GC M62. Finally, \citet{peuten16} used $N$-body star cluster simulations to demonstrate that the lack of mass segregation in NGC 6101 could be the result of the cluster having a larger stellar-mass BH population. 

Understanding the orbital behaviour of newly formed BHs and BH merger remnants that receive velocity kicks due to gravitational wave recoil at the centre of their host potential (which can either be a star cluster or a galaxy) is essential to understanding how gravitational waves are produced and how massive BHs may form and evolve. In this paper, we specifically explore the behaviour of kicked BHs in star cluster environments with masses that are comparable to the sources of all four gravitational wave detections. We also investigate how well classic dynamical friction theory \citep{chandrasekhar43} predicts the orbital decay of a kicked BH. Several studies have already attempted to model the evolution of a BH that has been given a velocity kick and is subject to dynamical friction, however they have been forced to restrict the parameter space to either near-zero velocity kicks \citep{chatterjee02} or high BH masses \citep{gualandris08} in order to develop an analytic model for how the BH's energy (and therefore its position and velocity) evolves with time. An analytic model to describe as a function of time the displacement from r $=$ 0 of a kicked massive BH, no matter the initial kick velocity or BH mass, will have direct implications for observations of BHs/IMBHs in GCs and NSCs, SMBHs in galactic nuclei, and gravitational wave detections. Modelling the orbital decay of kicked BHs will also help constrain theoretical models attempting to describe the formation, merger rate and subsequent growth of BHs. 

In Section~\ref{method}, we derive the equations of motion for a BH located at r $=$ 0 that is imparted with a kick at t $=$ 0, assuming a Plummer sphere for the gravitational potential. Section ~\ref{nbody} is then dedicated to introducing the $N$-body simulations we use to determine how well derived equations of motion predict the behaviour of kicked BHs in star clusters. We then compare our model, as well as the models of \citet{chatterjee02} and \citet{gualandris08} to our $N$-body simulations in Section~\ref{results}. In Section \ref{discussion} we address discrepancies between the theory and simulations and introduce a new formalism to properly model the evolution of BHs with $m_{bh} \lesssim 100 M_\odot$. The formalism is applicable over the entire range of BH masses and kick velocities explored in this study. Finally, we summarize our main conclusions and discuss the implications of our results in Section~\ref{summary}.  

\section{Theory}\label{method}

The initial mass function of star clusters is such that hundreds of BHs will form early in the cluster's lifetime \citep{salpeter55, kroupa93, kroupa01}. Natal velocity kicks between 0 and 100 km/s are such that approximately $10\%$ of all BHs are retained by the cluster \citep{pfahl02,pfahl03}. Over time, kicked BHs will sink to the core of the cluster through dynamical friction and two-body interactions. While some BHs will be subsequently kicked from the cluster due to BH-BH interactions, \citet{sippel13} found that just over $30\%$ of all retained BHs will remain in the cluster after a Hubble time as either single BHs, part of a binary system with a main sequence star, or part of a BH-BH binary.

Of the BHs that form BH-BH binaries, only a fraction will merge while still within the cluster while others will be kicked from the cluster due to three-body scattering before merging \citep{askar17,banerjee17a,banerjee17b}. \citet{askar17} suggests that the fraction of BH-BH mergers occurring within the cluster may be as low as $15\%$. As previously mentioned, when a BH-BH binary merges the velocity kick imparted on the merger product can be anywhere from 1 km/s to over 500 km/s. Hence, when mergers occur only a small fraction of BH-BH merger products will be retained by the cluster and will later form a new, more massive, BH-BH binary that will also eventually merge. Similar to newly formed BHs, dynamical friction and two-body interactions will force retained merger products to sink to the core again so the process can repeat. However, it should be noted that only a small fraction of all BHs produced in a cluster will reach the necessary masses to produce the gravitational wave detections produced by LIGO.

To better constrain the ability and timescale over which repeated BH-BH mergers will occur within star clusters, it is important to understand how effectively dynamical friction can return kicked BHs to the core. \citet{chandrasekhar43} was the first to derive the loss of energy via dynamical friction that a test body of mass M will experience due to a \textit{homogeneous} background distribution of stars of mass $m_s$. Given that the test body has a velocity $v_{M}$, the force of dynamical friction on the test body can be written as:

\begin{equation}
F_{DF} = - \beta v_{M}
\end{equation}.

where $\beta$ equals \citep{binney87}:

\begin{equation} \label{beta0}
\beta=16 \pi^2  ln\Lambda G^2 M m_s \frac{\int_0^{v_{M}} f(r,u) u^2 du}{v_{M}^3}
\end{equation}

In Equation \ref{beta0}, $ln \Lambda$ is the Coulomb logarithm and $f(r,u)$ is the phase space distribution function of stellar positions r and velocities u. However it should be noted that the calculations are just as easily applicable to a test body losing energy via dynamical friction due to a background gaseous field.

Several studies have considered the scenario of the test body travelling through a non-homogeneous stellar field, including cases where the test body passes through or is located at the minimum of a gravitational potential. Such a scenario directly applies to star clusters and satellite galaxies orbiting around a central host  \citep[e.g.][]{ostriker89, pesce92, colpi99, fujii06, cole12, arca14a, arca14b, brockamp14} and BHs orbiting within the background stellar field of either a star cluster or galaxy \citep[e.g.][]{chatterjee02, vicari07, gualandris08, antonini12}. In these cases, the mass of the test body is significantly higher than the mean mass of stars in the stellar field but only a fraction of the total mass of the stellar system.

Several of the above studies have found that the early work of \citet{chandrasekhar43} can be inaccurate when applied to test bodies in non-homogeneous stellar fields \citep{ostriker89, pesce92, colpi99, antonini12, arca14a, arca14b, vicari07}. Hence new treatments for the dynamical friction of test bodies in dense stellar systems have been developed to estimate in-fall times \citep{antonini12, arca14a}. Recently, the works of \citet{antonini12} and \citet{arca14a} have been demonstrated to be applicable to the inner regions of galaxies \citep{antonini14, arca15} and star clusters \citep{arca16} respectively.

We continue exploring the scenario of test bodies in non-homogeneous stellar fields by applying the work of \citet{chandrasekhar43} to newly formed BHs and BH-BH merger remnants orbiting within a star cluster environment. The resulting BH will have a mass m$_{\rm BH} \gg m$, where m is the mean stellar mass in the surrounding star cluster.  We assume the remnant is formed at time t $=$ 0 with a velocity kick of magnitude v$_{\rm kick}$ that is less than the escape velocity v$_{\rm esc}$ from r $=$ 0.

In order to obtain a solution for the displacement from r $=$ 0 or the position of the kicked BH r(t) as a function of time t, we begin by writing the equation of motion for the BH.  This is done by balancing the forces acting on the BH, beginning immediately after the BH receives a kick at t $=$ 0 and r $=$ 0:

\begin{equation}
\label{eqn:forces}
F_{\rm tot} = F_{\rm g} + F_{\rm DF},
\end{equation}

where F$_{\rm g}$ is the gravitational force and F$_{\rm DF}$ is the damping force, which we take as being equal to the stellar dynamical friction force F$_{\rm sDF}$ (note that we have absorbed the force direction into the terms F$_{\rm g}$ and F$_{\rm sDF}$; see below). Several studies have attempted to simplify Equation \ref{beta0} such that the exact form of the distribution function does not need to be known \citep[e.g.][]{chatterjee02,gualandris08}. The first key assumption that needs to be made moving forward is the functional form of the background potential. For the purposes of this study, we will assume the background stellar field can be approximated as a Plummer Sphere, which has been used to describe open clusters, GCs and galactic bulges.

For a Plummer sphere, the gravitational potential and density are, respectively:
\begin{equation}
\label{eqn:pot}
\Phi(r) = -\frac{GM}{\sqrt{r^2 + a^2}} = -\frac{GM}{a}\Big(1 + \frac{r^2}{a^2}\Big)^{-1/2}
\end{equation}
and
\begin{equation}
\label{eqn:dens}
\rho(r) = \frac{3M}{4{\pi}a^3}\Big(1 + \frac{r^2}{a^2}\Big)^{-5/2},
\end{equation}
where M is the total cluster mass, a is the Plummer radius or scale length and r is the distance from the cluster centre.  For a Plummer sphere, the velocity dispersion takes on its maximum value at r $=$ a, or:
\begin{equation}
\label{eqn:sigma}
\sigma(a) = \Big( \frac{GM}{2\sqrt{2}a} \Big)^{1/2}
\end{equation}

The specific gravitational force F$_{\rm g}$ acting on the kicked BH is calculated from the gradient of the gravitational potential:
\begin{equation}
\label{eqn:fg}
F_{\rm g} = -\frac{d\Phi}{dr} = \frac{GMr}{a^3}\Big(1 + \frac{r^2}{a^2}\Big)^{-3/2}
\end{equation}
Similarly, the specific stellar dynamical friction force acting on the BH is \citep{chandrasekhar43}:
\begin{equation}
\label{eqn:fdf}
F_{\rm sDF} = -\frac{4{\pi}G^2m_{\rm BH}{\rm ln\Lambda}\rho(r)}{v^2},
\end{equation}
where v $=$ $\dot{r} =$ dr/dt is the (magnitude of the) velocity of the BH with respect to the cluster centre, and $\ln \Lambda$ is the Coulomb logarithm. Note that we take $\ln \Lambda$ from \citet{chatterjee02}, who specifically derives the Coloump logarithm for a Plummer sphere. Plugging Equations~\ref{eqn:fg} and~\ref{eqn:fdf} into Equation~\ref{eqn:forces} gives:

\begin{equation}
\label{eqn:forces2}
\ddot{r} + \frac{3G^2Mm_{\rm BH}{\rm ln\Lambda}}{a^3}\Big(1 + \frac{r^2}{a^2}\Big)^{-5/2}\dot{r}^{-2} + \frac{GM}{a^3}\Big(1 + \frac{r^2}{a^2}\Big)^{-3/2}r = 0
\end{equation}

Note that we are implicitly assuming that only the BH moves in our model.  The cluster potential and 
centre of mass are assumed to be static in time and space.

Equation~\ref{eqn:forces2} is a second-order ordinary differential equation which must be solved numerically.  In order to obtain an analytic solution, simplifying assumptions must be made in addition to assuming that the stellar field can be approximated by a Plummer Sphere and that $m_{bh} \gg m_s$. For the purposes of this study, we will first consider the special cases of a BH following a 1D trajectory and:

\begin{itemize}
\item The BH is given a low initial velocity at the origin of the background potential such that it stays within the Plummer Sphere's scale radius.
\item The BH is undergoing Brownian motion near the origin of the background potential \citep{chatterjee02}.
\item  The BH is massive compared to nearby stars and is given a larger velocity kick such that its location within the background potential is not restricted \citep{gualandris08}.
\end{itemize}

In each section we describe the necessary assumptions and defer a more thorough discussion of their justification and astrophysical significance to Section~\ref{discussion}. 

\subsection{Low Velocity Kicks at the Origin}\label{sec:lvk}

By assuming that the motion of the kicked BH follows a 1D trajectory, we set $\dot{r} =$ 0 at the turn-around points.  This is not accounted for in Equation~\ref{eqn:fdf}; the dynamical friction force blows up at zero velocity.  To correct for this we take the simple, yet unphysical, approach of adding an additional term to the velocity in the denominator of Equation~\ref{eqn:fdf}, namely the product of the stellar velocity dispersion $\sigma$.  That is, we replace $\bf{\dot{r}}$v$^{-3}$ with $\bf{\dot{r}}$(v$^2$ + $\sigma^2$)$^{-3/2}$.  This gives:

\begin{equation}
\label{eqn:fdf2}
F_{\rm sDF} = -\frac{4{\pi}G^2m_{\rm BH}{\rm ln\Lambda}\rho(r)\dot{r}}{(\sigma^3)}\Big(1 + \frac{\dot{r}^2}{\sigma^2} \Big)^{-3/2}.
\end{equation}  

As required, Equation~\ref{eqn:fdf2} gives F$_{\rm sDF} =$ 0 when v $=$ 0, and is otherwise negative. For examples of more rigorous and dynamically motivated treatments for Equation~\ref{eqn:fdf} diverging as $\dot{r}$ goes to 0, see \citet{just11}, \citet{antonini12}, and \citet{arca14a}.

Next we assume that the BH is restricted to the limiting case v$_{\rm kick} \ll \sigma$ in order to get the last term in Equation~\ref{eqn:fdf2} to asymptote to unity. Under this assumption, we obtain:

\begin{equation}
\label{eqn:fdf3}
F_{\rm sDF} = -\frac{4{\pi}G^2m_{\rm BH}{\rm ln\Lambda}\rho(r)}{\sigma^3}\dot{r}
\end{equation}  

Similarly, if we assume that the motion of the kicked BH is restricted to r $\ll$ a, then Equation~\ref{eqn:fg} simplifies to:
\begin{equation}
\label{eqn:fg2}
F_{\rm g} = -\frac{d\Phi}{dr} = \frac{GM}{a^3}r
\end{equation}
Plugging Equations~\ref{eqn:fdf3} and~\ref{eqn:fg2} back into Equation~\ref{eqn:forces2} now yields an 
equation of the general form:
\begin{equation}
\label{eqn:forces3}
\ddot{r} - b\dot{r} - kr = 0,
\end{equation}

Equation~\ref{eqn:forces3} has a well known solution, namely that of the damped simple harmonic oscillator.  That is, in the limits $\dot{r} \le$ v$_{\rm kick} =$ $\dot{r}$(t $=$ 0) $\ll$ $\sigma$ and r $\ll$ a, the solution to Equation~\ref{eqn:forces2} is:

\begin{equation}
\label{eqn:r}
r(t) = Ae^{-bt/2}sin(\omega_{\rm DF}t),
\end{equation}

where the constants are 

\begin{equation}
\label{eqn:b}
b = \frac{3G^2m_{\rm BH}M{\rm ln\Lambda}}{\sigma^3a^3}
\end{equation}

and

\begin{equation}
\label{eqn:k}
k = \frac{GM}{a^3}
\end{equation}

The damping frequency is:

\begin{equation}
\label{eqn:omegad}
\omega_{\rm DF} = \omega_{\rm 0}\Big(1 - \frac{b^2}{4k} \Big)^{1/2},
\end{equation}

where $\omega_{\rm 0} = \sqrt{k}$ is the natural (undamped) frequency. Equation~\ref{eqn:r} is 
subject to the boundary conditions r(0) $=$ 0 and $\dot{r}$(0) $=$ v$_{\rm kick} =$ $\dot{r}_{\rm 0}$, 
which gives for the amplitude of oscillation:

\begin{equation}
\label{eqn:amp}
A = \frac{\dot{r}_{\rm 0}}{\omega_{\rm DF}}
\end{equation}

Equation~\ref{eqn:r} together with Equations~\ref{eqn:b}, ~\ref{eqn:k}, ~\ref{eqn:omegad} and~\ref{eqn:amp} give the position of the BH at any time t after receiving a kick at the origin, in the limit 
$\dot{r}_{\rm 0} =$ v$_{\rm kick} \ll \sigma$ and r $\ll$ a (i.e. for small BH kicks).  This should frequently be the case for mergers between stellar-mass BHs and either IMBHs or SMBHs at the centres of, respectively, massive GCs and galactic nuclei. 

\subsubsection{Over-Damped versus Under-Damped Oscillation}

Equation \ref{eqn:omegad} can also be used to qualify the BH's behaviour by comparing it to the \textit{BH's relaxation time} (Equation~\ref{eqn:taurhbh}), which corresponds roughly to the time required for the BH to reach its final kinetic energy, or (approximate) orbit within the cluster, at which point the rate of dynamical heating of the BH due to random perturbations with other stars is balanced by the rate of cooling due to dynamical friction.\footnote{When this equilibrium is reached, the final steady-state BH velocity should be v $\sim$ (m/m$_{\rm BH}$)$^{1/2}\sigma$, however recent studies have found the $\frac{1}{2}$ power is too high by a factor of $\sim$ 2 \citep{trenti13}.}  It can be calculated using Equation 3.2 in \citet{merritt13} at r $=$ a:

\begin{equation}
\label{eqn:taurhbh}
\tau_{\rm rh}(m_{\rm BH}) = \frac{m}{m_{\rm BH}}\tau_{\rm rh}(m),
\end{equation}
where
\begin{equation}
\label{eqn:taurh}
\tau_{\rm rh}(m) = 0.34\frac{\sigma(a)^3}{G^2m\rho(a){\rm ln\Lambda}},
\end{equation}
where $\rho$(a) and $\sigma$(a) are the Plummer density and velocity dispersion, respectively, evaluated at r $=$ a.  Equations~\ref{eqn:dens} and~\ref{eqn:sigma} can be plugged into Equation~\ref{eqn:taurh}, and subsequently into Equation~\ref{eqn:taurhbh}, which can then be re-written as:
\begin{equation}
\label{eqn:taurhbh2}
\tau_{\rm rh}(m_{\rm BH}) = \frac{0.91{\pi}M^{1/2}a^2}{G^{1/2}m_{\rm BH}{\rm ln\Lambda}}.
\end{equation}

We will henceforth refer to Equation~\ref{eqn:taurhbh2} as the \textit{BH relaxation time}. Equation~\ref{eqn:taurhbh2} roughly describes the time required for the kicked BH to become fully damped, to within a factor of $\sim$ 2.

\begin{figure}
\begin{center}
\includegraphics[width=\columnwidth]{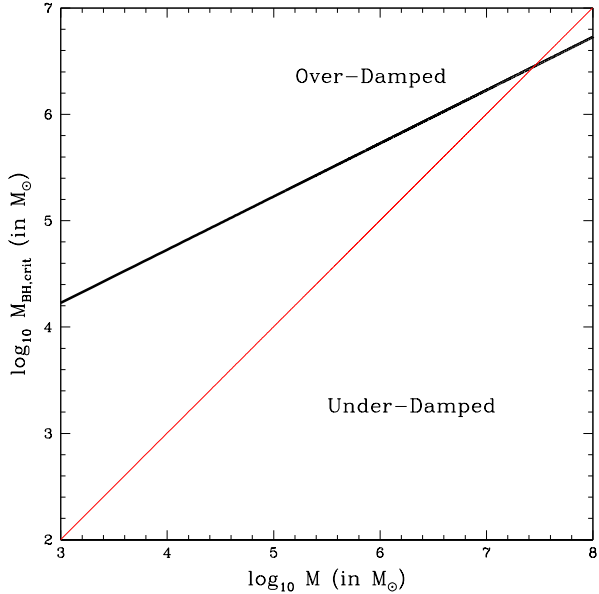}
\end{center}
\caption[The (logarithm of the) BH mass corresponding to the critically damped case as a function of the (logarithm of the) total cluster 
mass.]{The black line shows the (logarithm of the) BH mass corresponding to a critically damped oscillator, for which the BH is 
fully relaxed after only a single oscillation, as a function of the (logarithm of the) total cluster mass.  For comparison, 
the red line shows the BH mass for which the crossing time within the Plummer radius is equal to the BH 
relaxation time.  All masses are shown in solar masses.  
\label{fig:sho}}
\end{figure}

In Figure~\ref{fig:sho}, we show by the solid line the critical BH mass m$_{\rm BH}$ at which the oscillator is critically damped, as a function of the total cluster mass M. That is, the solid black line corresponds to the "critical" BH mass at which the BH's relaxation time is equal to the period of oscillation, 2$\pi$/$\omega_{\rm DF}$. Above this line the cluster's relaxation time is shorter than the BH's orbital period, such that the cluster's internal dynamics will return the BH to the origin very quickly and it will behave as an over-damped oscillator. For a BH which falls below the line, it will be able to undergo several complete orbits and have its orbit decay due to dynamical friction before the cluster's internal dynamics begin to play an important role. Hence it will behave as an under-damped oscillator. For comparison we also show by the red line the critical BH mass at which the crossing time, which can also be used as a tracer of a cluster's dynamical age, inside the Plummer radius is equal to the BH relaxation time. Importantly, Figure~\ref{fig:sho} is valid only for small kick velocities satisfying v$_{\rm kick} < \sigma$.  

\subsection{Brownian Motion}\label{sec:chat}

\citet{chatterjee02} applied the work of \citet{chandrasekhar43} to the dynamics of a MBH undergoing Brownian motion near the center of a dense stellar system characterized by a Plummer sphere of mass $M$ and scale radius $a$. Since the distribution function varies slowly with r near the center of a Plummer sphere, the authors also used Equation \ref{beta0} under the assumption that $r_{bh} << a$ at all times. Hence the background potential can be approximated to be $\Phi = - \frac{G M_p}{a}$. Furthermore, since the authors were not focused on the BH undergoing oscillatory motion, they did not have to adjust their derivation to take into account $F_{DF}$ going to infinity at zero velocity as was done in Equation \ref{eqn:b}. 

Assuming the MBH moves very slowly compared to nearby stars,  \citet{chatterjee02} instead reduces Equation \ref{beta0} to:

\begin{equation} \label{beta_chat}
\beta=\frac{128 \sqrt{2}}{7\pi} ln\Lambda (\frac{G}{M a^3})^{\frac{1}{2}} m_{bh}
\end{equation}

\noindent The authors then derive an equation of motion for a BH given an initial position and velocity similar to Equation \ref{eqn:r}, with the exception that their definition of $\beta$ differs slightly from Equation \ref{eqn:b} and they do not require the BH to be located at the origin at time zero.

\subsection{Large Velocity Kicks} 

\citet{gualandris08} also applied the work of \citet{chandrasekhar43} to BH dynamics, however instead focussed on the ejections of SMBHs from galaxy cores. Hence the properties of both the BH and the background stellar field considered by \citet{gualandris08} are different than those addressed in Sections \ref{sec:lvk} and \ref{sec:chat}. The main difference being that \citet{gualandris08} do not restrict the location of their BH to the inner regions of a stellar field. 

Using only the assumption that $m_{bh} >> m_s$,  \citet{gualandris08} write $\beta$ as:

\begin{equation} \label{beta_gual}
\beta=2 \pi G^2 \rho(r_{bh}) m_{bh} ln (1+\Lambda^2) v_{bh}^{-3} N(<v_{bh},r)
\end{equation}

where $\rho(r)$ is the mass density of stars at the BH's position and $N(<v_{bh},r)$ is the fraction of stars at the BH's position that are moving with velocities less than $v_{bh}$ (in the frame of the galaxy). Since $\beta$ in Equation \ref{beta_gual} is a function of r, the BH's equation of motion cannot be simply set equal to that of the damped oscillator. Hence \citet{gualandris08} compared the orbital decay of kicked BHs over short time steps in $N$-body simulations to the decay predicted by $F_{DF}=-\beta v_{bh}$ and Equation \ref{beta_gual}. The authors found that the work of \citet{chandrasekhar43} could initially reproduce the orbital decay of the kicked BH (for $2 < ln \Lambda < 3$) as long as the evolution of $\rho(r_{bh})$ was accounted for up until when the amplitude of motion falls below the background potential's core radius. Afterwards, \citet{gualandris08} note that the SMBH and the core oscillate about their center of mass for a long period of time until the oscillations damp to the Brownian level. 

Our approach, along with the studies of \citet{gualandris08} and \citet{chatterjee02}, determine the equation of motion of the BH by making assumptions regarding the mass of the BH (e.g. $m_{bh} >> m_{star}$), the kick velocity ($v_{k} << \sigma_{star}$), and its location within the cluster ($r << a$) in order to reach an analytic solution. However this significantly limits the BH mass - kick velocity parameter space and leaves many combinations unexplored. Using a suite of $N$-body models, we will compare each of these approaches to simulations of BHs evolving in star clusters over a range of BH masses and initial kick velocities to identify the regions of parameter space that each study can successfully reproduce and identify the regions of parameter space that need further consideration.

\section{$N$-body Simulations} \label{nbody}

To model the evolution of kicked BHs in star cluster environments, we use the direct $N$-body code NBODY6 \citep{aarseth03}. Each star cluster is initially a Plummer Sphere of 50,000 stars with an initial half-mass radius $r_m$ of 2.5 pc.  To isolate and identify the effects of a non-homogeneous stellar field on the orbital evolution of BHs, we assume an equal mass cluster where all stars are 0.5 $M_\odot$. Hence model clusters have velocity dispersions and density profiles comparable to open clusters. The kicked BH in each model has a mass of either 10, 50, or 100 $M_\odot$ and starts at the center of the cluster with a kick velocity of 3.6, 4.8, 6.0, 7.2, 8.4 or 9.6 km/s. The velocities correspond to $\frac{v_{k}}{\sigma(a)}$ values of 0.78, 1.0, 1.3, 1.6, 1.8, and 2.0 and are such that the BH does not  escape the cluster.

\begin{figure}
\begin{center}
\includegraphics[width=\columnwidth]{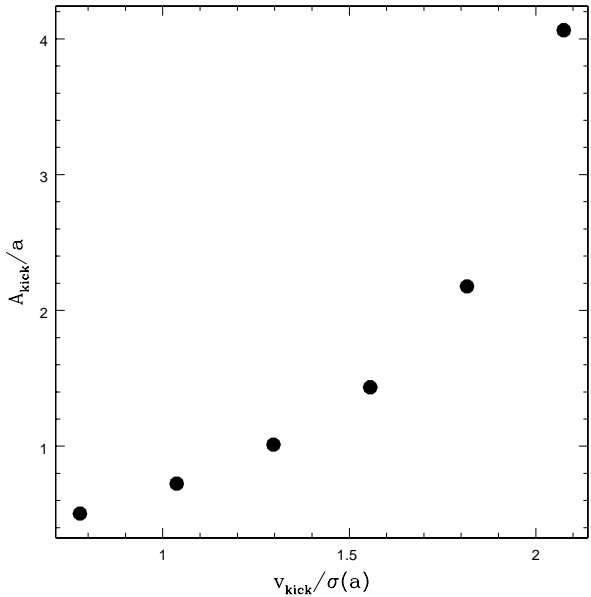}
\end{center}
\caption[Ratio of initial kick velocity to velocity dispersion at the Plummer scale radius versus the ratio of initial orbital amplitude to the Plummer scale radius for simulations with a range of initial kick velocities (marked in the legend).]{Ratio of initial kick velocity to velocity dispersion at the Plummer scale radius versus the ratio of initial orbital amplitude to the Plummer scale radius for simulations with a range of initial kick velocities (marked in the legend).
\label{fig:vamp}}
\end{figure}

With respect to Figure \ref{fig:sho}, each of these kicked BHs will behave as under-damped oscillators. Figure \ref{fig:vamp} illustrates the range in $\frac{v_{k}}{\sigma(a)}$ and $\frac{A_{kick}}{a}$ covered by our simulations, where $A_{kick}$ is the orbital amplitude associated with $v_{k}$ calculated explicitly assuming energy is conserved during the time it takes for the BH to travel from the origin to its maximum clustercentric distance. 

\begin{figure}
\begin{center}
\includegraphics[width=\columnwidth]{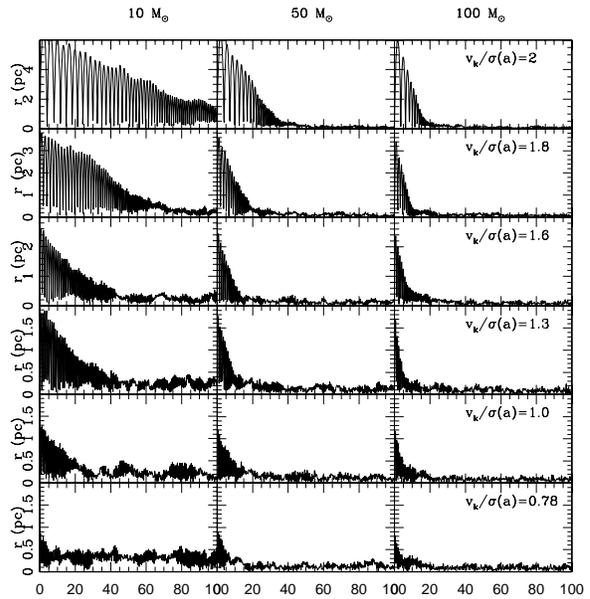}
\end{center}
\caption[Orbital evolution of kicked BHs with masses of 10 $M_\odot$ (left column), 50 $M_\odot$ (center column) and 100 $M_\odot$ (right column). Different rows correspond to different values of $\frac{v_{k}}{\sigma(a)}$, which are marked in the legend.]{Orbital evolution of kicked BHs with masses of 10 $M_\odot$ (left column), 50 $M_\odot$ (center column) and 100 $M_\odot$ (right column). Different rows correspond to different values of $\frac{v_{k}}{\sigma(a)}$, which are marked in the legend.
\label{fig:ramp}}
\end{figure}

For illustrative purposes, we have plotted the orbital evolution of the BH in each simulation in Figure \ref{fig:ramp}. In agreement with previous studies, a kicked BH acts as a damped oscillator while its orbit decays due to dynamical friction. The orbits of BHs given low velocity kicks decay much quicker than BHs that are given large velocity kicks, entering the Brownian motion phase at much earlier times. Additionally, consistent with Equation \ref{beta0}, higher mass BHs decay faster than lower mass BHs.  

\section{Comparing Theory to Simulations}\label{results}

\begin{figure}
\begin{center}
\includegraphics[width=\columnwidth]{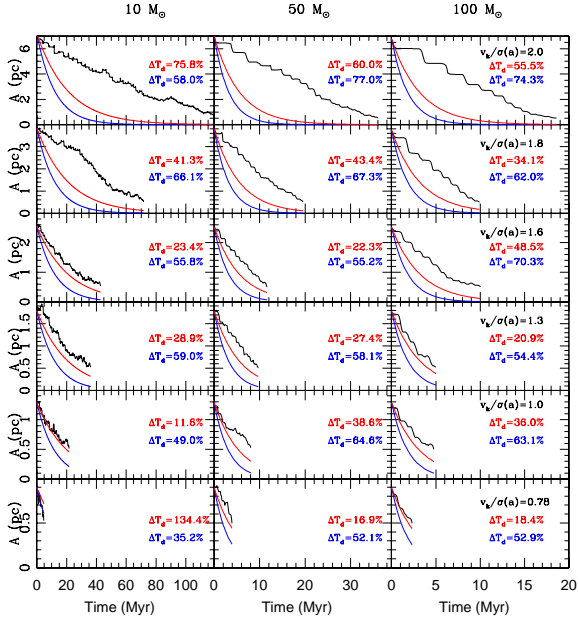}
\end{center}
\caption[Orbital amplitude of kicked BHs with masses of 10 $M_\odot$ (left column), 50 $M_\odot$ (center column) and 100 $M_\odot$ (right column) as a function of time. Different rows correspond to different values of $\frac{v_{k}}{\sigma(a)}$, which are marked in the legend. The models are compared to the predictions of Equation \ref{eqn:b} (blue) and Equation \ref{beta_chat} (red), with the percent difference between the actual and expected decay times noted in each panel.]{Orbital amplitude of kicked BHs with masses of 10 $M_\odot$ (left column), 50 $M_\odot$ (center column) and 100 $M_\odot$ (right column) as a function of time. Different rows correspond to different values of $\frac{v_{k}}{\sigma(a)}$, which are marked in the legend. The models are compared to the predictions of Equation \ref{eqn:b} (blue) and Equation \ref{beta_chat} (red), with the percent difference between the actual and expected decay times noted in each panel. 
\label{fig:de}}
\end{figure}

In order to compare our $N$-body simulations to the theoretical predictions discussed in Section \ref{method}, we first compare the decay of the BH's orbital amplitude in each simulation to the predicted decay rates of Equation \ref{eqn:b} (blue) and Equation \ref{beta_chat} (red) from \citet{chatterjee02}. This comparison is made in Figure \ref{fig:de}, with the the percent difference between the actual and expected decay times noted in each panel. It should be noted that we only plot the BH's decay up to the point that it enters the Brownian motion phase. Once the BH has settled within the core of the cluster and starts undergoing brownian motion, the theoretical predictions are not expected to be representative of the changes in energy experienced by the BHs in our models \citep{gualandris08}.

From Figure \ref{fig:de}, we see that Equations \ref{eqn:b} and \ref{beta_chat} significantly overestimate how much energy the BH loses via dynamical friction as they yield decay rates much higher than observed in the models. Specifically, both equations predict a much higher energy loss rate when the BH is near its turnaround point in the cluster's core. The discrepancies can be attributed to the fact that both equations assume $v_{k} << \sigma$ and $r << a$, which is not the case here, in order to derive an equation of motion for the BH. Even for the lower kick velocity cases, which are closest to the theoretical estimates, only the lowest mass model agrees with dynamical friction theory. The other models are still in disagreement with theoretical estimates because the initial assumption made by \citet{chandrasekhar43} that the background stellar field is homogeneous is also not applicable here. 

An additional factor that none of the equations account for is energy gains by the BH due to two-body interactions. With the exception of the $\frac{v_{k}}{\sigma(a)} = 0.78$ models, the kicks received by the BHs result in their decay times being longer than the BH's relaxation time within the cluster. Hence two-body interactions are starting to affect the BHs orbital evolution just as much as, if not more than, dynamical friction. When the BH passes through the dense environment of the cluster's core, the local potential can vary significantly. When this occurs, major episodes of energy loss and/or gain can occur due to close encounters between the BH and nearby stars. At later times, when the BH's orbital velocity has decreased it will also be affected by interactions with stars that are travelling faster than the BH itself \citep{antonini12, arca14a, dosopoulou17}.

Since $\beta$ in Equation \ref{beta_gual}  is a function of r, a decay rate cannot be determined using the approach of \citet{gualandris08}. Hence to compare our $N$-body simulations to \citet{gualandris08}, we instead determine the change in the BH's energy at each time step and compare it the specific energy loss via the force of dynamical friction acting on the BH. We calculate the specific energy loss predicted by each theory as $\Delta E = - \beta v_{bh} d$, where $\beta$ is taken from Equation \ref{beta_gual} and d is the distance travelled by the BH between time steps. Hence we are assuming $F_{DF}$ is constant between time steps. 

\begin{figure}
\begin{center}
\includegraphics[width=\columnwidth]{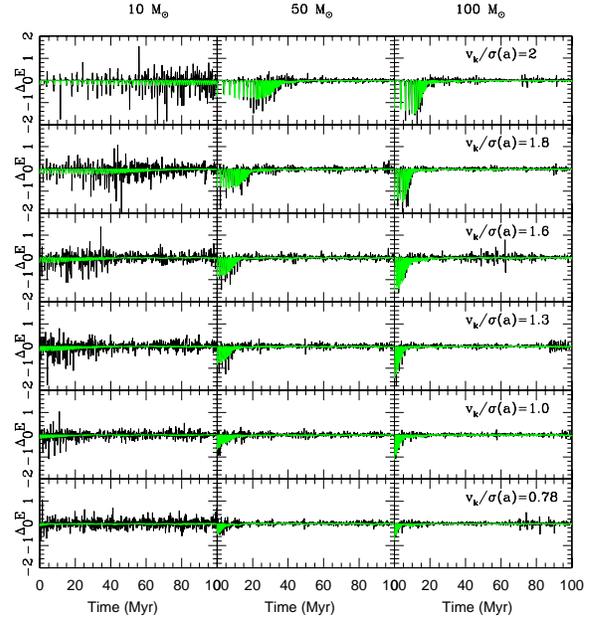}
\end{center}
\caption[Change in specific energy of kicked BHs with masses of 10 $M_\odot$ (left column), 50 $M_\odot$ (middle column) and 100 $M_\odot$ (right column) as a function of time. Different rows correspond to different values of $\frac{v_{k}}{\sigma(a)}$, which are marked in the legend. The green line illustrates the predicted change in energy by \citet{gualandris08}.]{Change in specific energy of kicked BHs with masses of 10 $M_\odot$ (left column), 50 $M_\odot$ (middle column) and 100 $M_\odot$ (right column) as a function of time. Different rows correspond to different values of $\frac{v_{k}}{\sigma(a)}$, which are marked in the legend. The green line illustrates the predicted change in energy by \citet{gualandris08}.
\label{fig:demerr}}
\end{figure}

From the change in energy predicted by Equation \ref{beta_gual} in Figure \ref{fig:demerr}, we see there are significant discrepancies over the course of the BHs decay. In fact, Equation \ref{beta_gual} typically underestimates the amount of energy lost by the BH as it passes through the core. This is likely due to the fact that Equation \ref{beta_gual} assumes $m_{bh} >> m_s$ (which again is not the case here), since \citet{gualandris08} does an increasingly better job of matching the simulations as $m_{bh}$ is increased. Furthermore, as previously stated, the fact that the BH sometimes gains energy while passing through the inner regions of the cluster is not taken into account by the works of \citet{chandrasekhar43}, which \citet{gualandris08} is based on. With \citet{chandrasekhar43} being unable to model changes in the local cluster potential, and the assumptions made by various works leading to both under and over estimating the amount of energy lost by the BH when it passes through denser environments, it appears that the only way to truly dynamically model the effects of dynamical friction on BH evolution over a wide range of BH masses and velocity kicks is knowing the exact form of the stellar distribution function at all times. 

\section{Discussion}\label{discussion}

\subsection{Two-Body Interactions versus Dynamical Friction}

As observed in Figures \ref{fig:de} and \ref{fig:demerr}, classic dynamical friction has trouble predicting the energy change experienced by low-mass BHs passing through the dense core of a stellar population. Equation \ref{eqn:b} and Equation \ref{beta_chat} both overestimate energy loss due to dynamical friction in the core while the \citet{gualandris08} model underestimates energy loss. None of the models account for energy gains by the BH due to two-body interactions, which is an important factor in models that have decay times longer than the BH relaxation time. Two-body interactions are specifically important when the local potential, and fluctuations thereof, experienced by the BH is more dominant than the force due to dynamical friction acting on the BH and when the BH encounters stars with $v>v_{bh}$ \citep{antonini12, arca14a, dosopoulou17}.

\begin{figure}
\begin{center}
\includegraphics[width=\columnwidth]{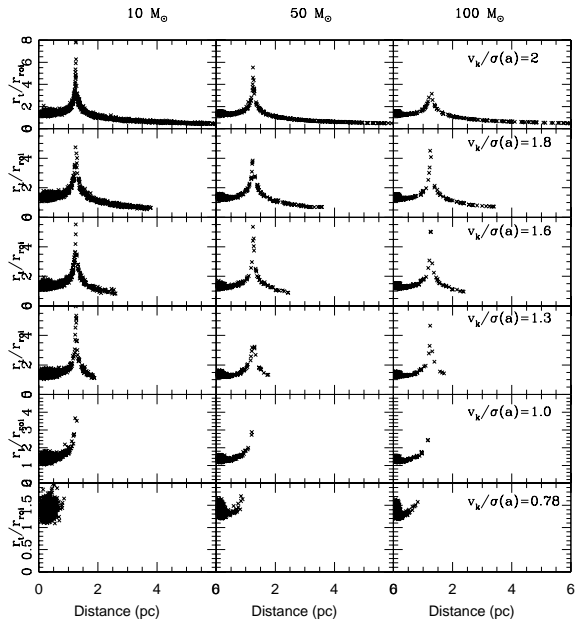}
\end{center}
\caption[Ratio of BH tidal radius to radius of sphere of influence as a function of clustercentric distance for kicked BHs with masses of 10 $M_\odot$ (left column), 50 $M_\odot$ (middle column) and 100 $M_\odot$ (right column) as a function of time. Different rows correspond to different values of $\frac{v_{k}}{\sigma(a)}$, which are marked in the legend.]{Ratio of BH tidal radius to radius of sphere of influence as a function of clustercentric distance for kicked BHs with masses of 10 $M_\odot$ (left column), 50 $M_\odot$ (middle column) and 100 $M_\odot$ (right column) as a function of time. Different rows correspond to different values of $\frac{v_{k}}{\sigma(a)}$, which are marked in the legend.
\label{fig:rrat}}
\end{figure}

To illustrate where in the star cluster each mechanism dominates, we plot the ratio of the BH's instantaneous tidal radius $r_t$ to the instantaneous radius of its sphere of influence $r_{soi}$ as a function of clustercentric distance in Figure \ref{fig:rrat}. The BH's sphere of influence represents a sphere within which the total mass of stars is equal to the mass of the BH. $r_t$ is calculated analytically based on the potentials of both the BH and the cluster ($r_t = (\frac{2.0 m_{bH}}{d^2\Phi(r)/dr^2})^{\frac{1}{3}}$). When $r_{soi}$ is larger than $r_t$, dynamical friction will be the dominant force acting on the BH as nearby stars are able to respond to the BH's passing and form a wake behind it. However when $r_{soi}$ is less than $r_t$, both the motion of the BH and nearby stars are primarily affected by both the strength of and variations in the local potential. Not only will the affects of dynamical friction be minimized, but variations in the local potential can result in the BH gaining or losing energy depending on the direction of its motion and the direction of the net force acting on the BH.

Figure \ref{fig:rrat} illustrates that in each case $\frac{r_t}{r_{soi}}$ sharply increases near 1.25 pc when the cluster passes in and out of the core, meaning two-body interactions become the dominant source of energy loss/gain for the BH. For the BH masses and kick velocities considered here, when the BH is within 1.25 pc of the cluster's center it is not surprising that classical dynamical friction will break down. For the lower kick velocities that Equations \ref{eqn:b} and \ref{beta_chat} are designed for, since the BHs kinetic energy will be similar to that of background stars the effects of two-body interactions will be minimized. For the case of $m_{bh} >> 100 M_\odot$ that \citet{gualandris08} was designed for, the sphere of influence would be much larger and $\frac{r_t}{r_{soi}}$ would not rise above 1.

\subsection {Correcting Orbit Decay}\label{sec:c}

In order to more accurately predict the energy evolution of a kicked BH, we aim to develop either a correction factor or a new formalism that accounts for the assumptions made by classic dynamical friction theory and two-body interactions experienced by the BH within the core of the cluster. Since the true decay rate is less than predicted by Equation \ref{eqn:b}, we first introduce a correction factor c such that the true decay rate is $\frac{b}{c}$. The parameter c represents a free parameter that we can calculated from simulations in order to compensate for the assumptions of \citet{chandrasekhar43} and the additional assumptions we have made in order to derive an analytical description of the BH's orbital decay (e.g. $v_{k} << \sigma_{star}$ and $r << a$). We have elected to correct Equation \ref{eqn:b} over Equation \ref{beta_chat} due to its treatment of energy lost by the black hole at the turnaround points of its orbit.

To determine the correction factor, we first assume the orbital decay of each BH model in Figure \ref{fig:de} can be treated as an exponential decay of the form

\begin{equation}\label{eqn:ampe}
A(t)=A_0e^{-bt/2c}
\end{equation}

where b is taken from Equation \ref{eqn:b}. We also consider the case where a BH's amplitude decays linearly instead of exponentially via:

\begin{equation}\label{eqn:ampl}
A(t)=\frac{c_L t}{2} + A_0
\end{equation}

\begin{figure*}
\begin{center}
\includegraphics[width=\textwidth]{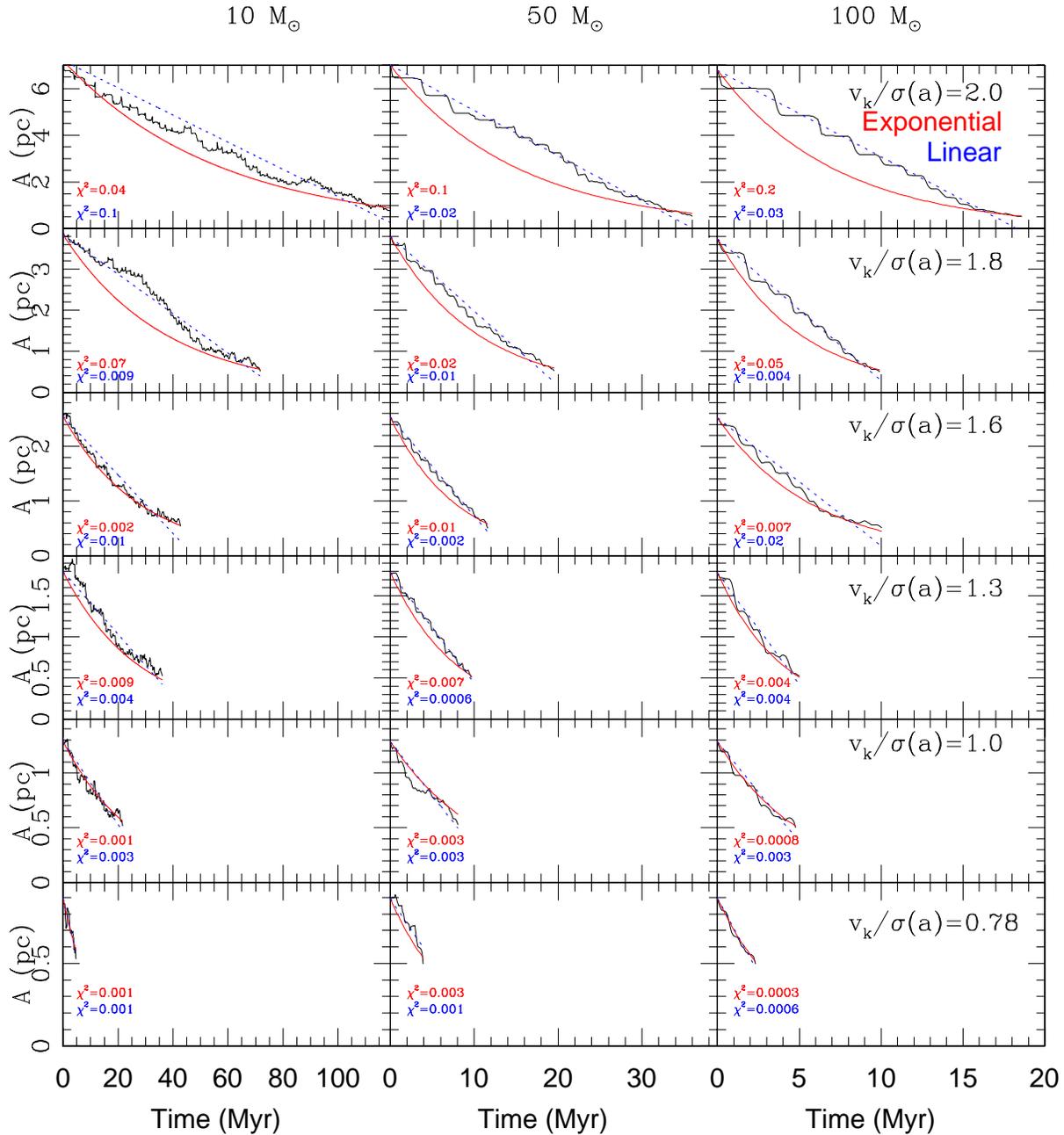}
\end{center}
\caption[Orbital amplitude of kicked BHs with masses of 10 $M_\odot$ (left column), 50 $M_\odot$ (center column) and 100 $M_\odot$ (right column) as a function of time. Different rows correspond to different values of $\frac{v_{k}}{\sigma(a)}$, which are marked in the legend. The models are compared to the predictions of Equation \ref{eqn:ampe} (red) and Equation \ref{eqn:ampl} (blue), with the corresponding reduced $\chi^2$ value noted in each panel.]{Orbital amplitude of kicked BHs with masses of 10 $M_\odot$ (left column), 50 $M_\odot$ (center column) and 100 $M_\odot$ (right column) as a function of time. Different rows correspond to different values of $\frac{v_{k}}{\sigma(a)}$, which are marked in the legend. The models are compared to the predictions of Equation \ref{eqn:ampe} (red) and Equation \ref{eqn:ampl} (blue), with the corresponding reduced $\chi^2$ value noted in each panel. 
\label{fig:beta}}
\end{figure*}

The orbital decay of each BH simulation and the best fit exponential and linear decay models (found using least squares fitting) are illustrated in Figure \ref{fig:beta}. The reduced $\chi^2$ values between each fit and the simulated data are noted in each panel.

From Figure \ref{fig:beta} it can be seen that for low and intermediate kick velocities ($\frac{v_{k}}{\sigma} \le 1.6$), the functional form of the decay is still exponential albeit with a decay rate that is much lower than Equation \ref{eqn:b} predicts. The simplified assumption of replacing 
$\bf{\dot{r}}$v$^{-3}$ with $\bf{\dot{r}}$(v$^2$ + $\sigma^2$)$^{-3/2}$ in order to calculate $F_{\rm sDF}$ when $\dot{r}=0$ is likely a contributing factor to this discrepancy. However, for the majority of cases the linear decay yields a better fit to the simulations (especially for higher mass BHs). In fact, for larger $\frac{v_{k}}{\sigma}$ ratios the exponential decay approach completely breaks down marking a clear departure from classic dynamical friction theory. Hence only the linear decay formalism is applicable over the entire range of $\frac{v_{k}}{\sigma}$ presented here.

\begin{figure}
\begin{center}
\includegraphics[width=\columnwidth]{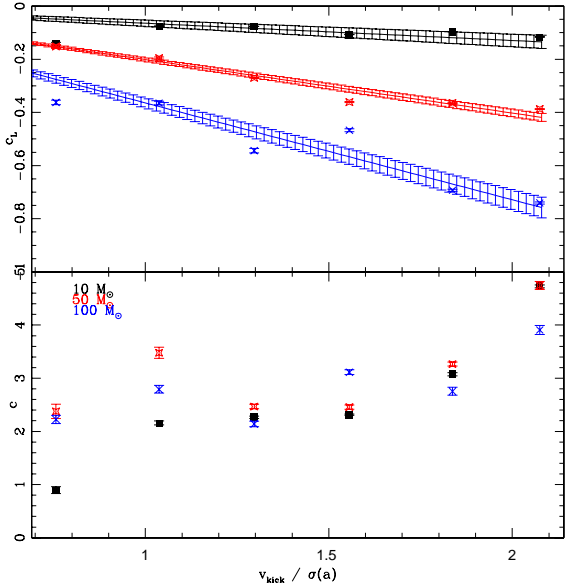}
\end{center}
\caption[Correction factor $c$ (lower panel) and $c_L$ (upper panel) as a function of $\frac{v_{k}}{\sigma(a)}$ for BHs with masses of 10 $M_\odot$ (black squares), 50 $M_\odot$ (red stars) and 100 $M_\odot$ (blue crosses). In the upper panel, solid coloured lines represent a linear fit to $c_L$ versus $v_{k}$ for each BH mass.] {Correction factor $c$ (lower panel) and $c_L$ (upper panel) as a function of $\frac{v_{k}}{\sigma(a)}$ for BHs with masses of 10 $M_\odot$ (black squares), 50 $M_\odot$ (red stars) and 100 $M_\odot$ (blue crosses). In the upper panel, solid coloured lines represent a linear fit to $c_L$ versus $v_{k}$ for each BH mass.
\label{fig:brat}}
\end{figure}

Using the best fit decay rates to solve for the correction factors $c$  and $c_L$ in each case, we plot the relationship between both factors and $\frac{v_{k}}{\sigma(a)}$ for all three BH masses in Figure \ref{fig:brat}. The uncertainty in each fitted decay rate is also plotted in Figure \ref{fig:brat}.

The lower panel of Figure \ref{fig:brat} demonstrates that for $\frac{v_{k}}{\sigma} < 1.6$, where the orbital decay is exponential, the decay constant b in Equation \ref{eqn:b} needs to be decreased by a factor between one and three. The trend appears to be primarily dependent on kick velocity and independent of BH mass. However there is significant scatter about any function that attempts to relate $c$ to $\frac{v_{k}}{\sigma}$, likely due to the effects that random close encounters have on the BH's decay rate. For example, a power-law fit to the data has an uncertainty greater than $80\%$.
 
It should also be noted that for the case where $m_{bh} = 10 M_\odot$ and $v_{k} = 3.5$ km/s, the decay is already well fit by an uncorrected decay constant ($c \sim 1$) as both the kick velocity is low and the BH never travels beyond the scale radius of the Plummer sphere. The BH's low mass also means that it is susceptible to two body interactions. Hence it is more accurate to consider the $m_{bh} = 10 M_\odot$,  $\frac{v_{k}}{\sigma(a)}=0.78$ model to be in the brownian motion phase at time zero.



The upper panel of Figure \ref{fig:brat} illustrates that higher kick velocities ($\frac{v_{k}}{\sigma} > 1.6$) result in faster mean linear decay rates. For a given BH mass, $c_L$ scales linearly with $\frac{v_{k}}{\sigma}$ over the entire range covered in the simulations as illustrated by the lines of best fit to each dataset. It is important to note that for each fit we have forced the fits to go through the origin. We also find that the slope of each line of best fit scales linearly with BH mass, such that $c_L$ can be written as:

\begin{equation}\label{clfit}
c_L= \Big((-0.003 \pm 4.0E-5) *m_{bh}-(0.03 \pm 0.003)\Big)  \times \frac{v_{k}}{\sigma}
\end{equation}

\begin{figure*}
\begin{center}
\includegraphics[width=\textwidth]{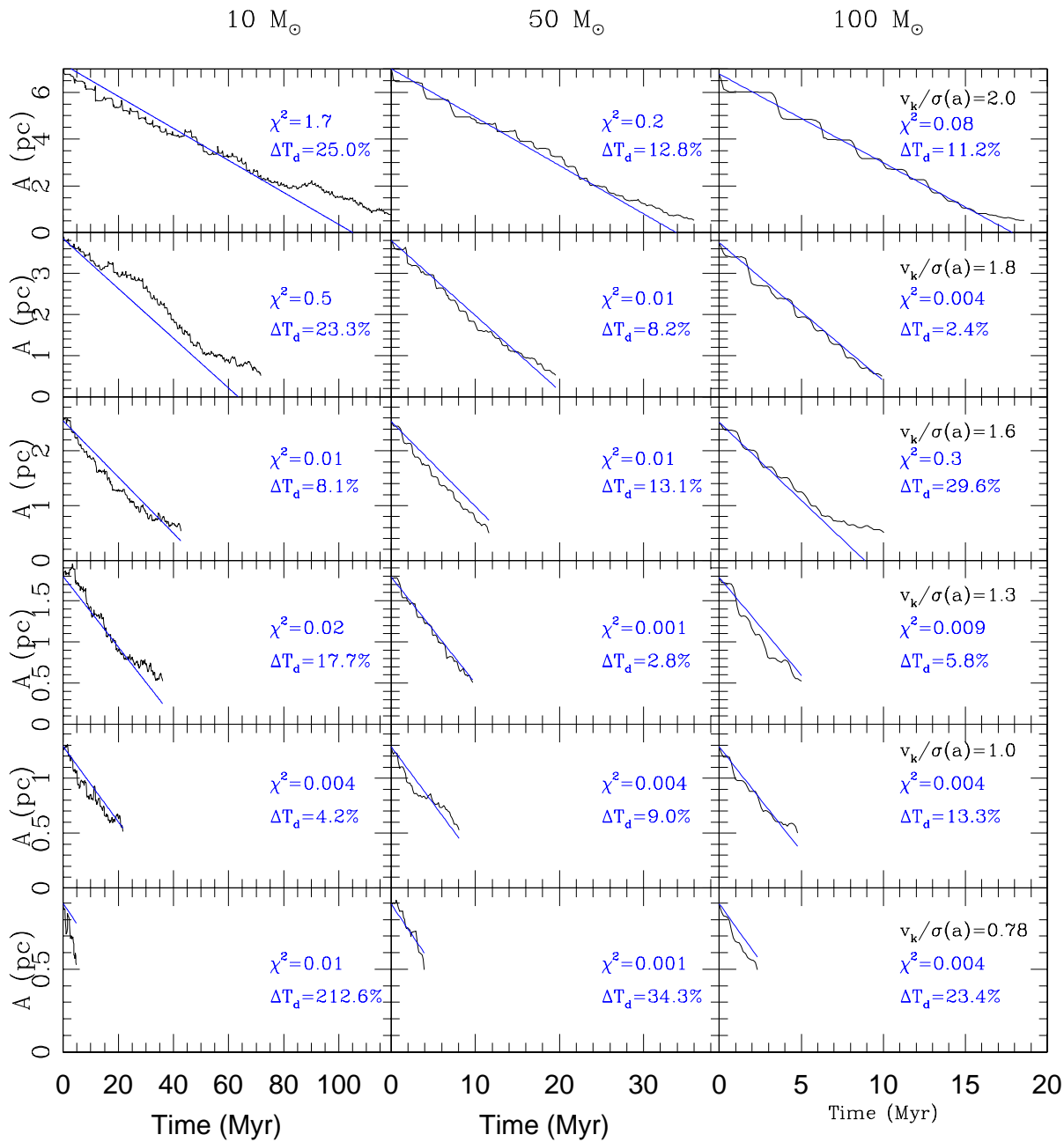}\end{center}
\caption[Orbital amplitude of kicked BHs with masses of 10 $M_\odot$ (left column), 50 $M_\odot$ (center column) and 100 $M_\odot$ (right column) as a function of time. Different rows correspond to different values of $\frac{v_{k}}{\sigma(a)}$, which are marked in the legend. The models are compared to the predictions of Equation \ref{eqn:ampl} with $c_L$ taken from Equation \ref{clfit} (blue), with the corresponding reduced $\chi^2$ and percent difference between the actual and expected decay times noted in each panel.]{Orbital amplitude of kicked BHs with masses of 10 $M_\odot$ (left column), 50 $M_\odot$ (center column) and 100 $M_\odot$ (right column) as a function of time. Different rows correspond to different values of $\frac{v_{k}}{\sigma(a)}$, which are marked in the legend. The models are compared to the predictions of Equation \ref{eqn:ampl} with $c_L$ taken from Equation \ref{clfit} (blue), with the corresponding reduced $\chi^2$ value and percent difference between the actual and expected decay times noted in each panel. 
\label{fig:fit}}
\end{figure*}

Taking into consideration that the majority of the simulations are better fit by a linear decay rate, and noting that even when an exponential decay is preferred a linear decay still yields a comparable reduced $\chi^2$, it appears that the orbital decay of kicked BHs with $m_{bh} < 100$ and $\frac{v_{k}}{\sigma} < 2.0$ is best modelled by Equations \ref{eqn:ampl} and \ref{clfit}. To illustrate the effectiveness of using our correction factor $c_L$, we plot the corrected theoretical orbital decay of each BH in Figure \ref{fig:fit} with the reduced $\chi^2$ value comparing the model to the simulations and the percent difference between the actual and expected decay times noted in each panel.

With the exception of the $m_{bh}=10 M_\odot$ - $\frac{v_{k}}{\sigma(a)}=0.78$ model, Figure \ref{fig:fit} illustrates that the linear decay formalism accurately reproduces the orbital decay of a kicked BH. In fact, directly comparing the $\Delta T_{d}$ values in Figure \ref{fig:fit} to Figure \ref{fig:de} illustrates that the linear formalism marks a significant improvement over Equations \ref{eqn:b} (blue) and Equations \ref{beta_chat} (red) for models with $\frac{v_{k}}{\sigma(a)}>0.78$. For models with $\frac{v_{k}}{\sigma(a)}<0.78$, the assumptions made when deriving Equations \ref{eqn:b} and \ref{beta_chat} ($v_{k} < \sigma$, $r < a$) still hold such that either equation can be applied to the lowest kick velocity models. Furthermore, since the decay time is less than the BH relaxation time in the low $v_k$ models, the effects of two-body interactions are minimized. However, some discrepancies between the linear formalism and the $\frac{v_{k}}{\sigma(a)}>0.78$ models remain in select cases when there is a sharp increase or decrease in energy or when the BH starts approaching the Brownian motion phase. In these cases, two-body interactions have become the dominant mechanism behind a BH's orbital evolution. Hence a departure from our dynamical friction model is not surprising.

\subsection{Model Limitations}

The $N$-body models presented here have been setup to specifically probe how classical DF theory handles the orbital decay of a kicked BH in a non-homogenous stellar field. The size and mass of our models, selected to optimize computation time, are such that they most directly reflect the evolution of a kicked BH in an open cluster. The applicability of our models to GC and NSCs will depend on multiple factors, mainly the effects of a complete mass spectrum, the total mass and density profile of the cluster, and in the case of NSCs the presence of a central SMBH.

Including a mass spectrum results in the production of an entire BH sub-population and allows for BHs to interact with stars over a range of stellar masses. Interactions with other BHs can cause the in-falling BHs to be kicked from a cluster entirely \citep{sippel13}, while the presence of a central SMBH has been shown to strongly affect the in-fall process \citep{antonini12, arca14a}. Our single-mass model clusters are most comparable to older stellar clusters when the effects of stellar evolution are minimal, other BHs have been kicked from the cluster, and only a strong binary remains \citep{sippel13, spera16}.

The total mass of our $N$-body star clusters $M_{SC}$, the main driver behind the computational time of our models, also limits the ranges of $\frac{m_{bh}}{M_{SC}}$ and $\frac{v_{k}}{\sigma}$ that we can directly probe as BHs with larger kick velocities will immediately escape the model clusters. Since our models are based on simulations of BHs in stellar fields that are comparable in size and mass to open clusters, additional simulations are required to determine whether the correction factors are scalable to the environments of GCs and NSCs as well. The linear dependence of $c_L$ on $m_{bh}$ and $\frac{v_{k}}{\sigma}$ suggests scaling to larger values of $\frac{v_{k}}{\sigma}$ may be possible. However with the linear formalism starting to break down for lower $m_{bh}$, how our models scale as $\frac{m_{bh}}{M_{SC}}$ decreases requires further study. 

\section{Conclusion} \label{summary}

We have used direct N-body simulations of BHs in star clusters to investigate how well classical DF theory predicts the orbital decay of BHs located at the center of a potential well that have received a velocity kick either at formation or due to the anisotropic emission of gravitational waves associated with a BH-BH merger. We specifically focus on BHs with $m_{bh} < 100$, which corresponds to the estimated merger products of the the four gravitational wave detections to date \citep{abbott16, abbott16b, abbott17, abbott17b}. Our models indicate that, over the range in BH mass and velocity kick explored here, classical dynamic theory only applies to BHs that receive very small velocity kicks ($\frac{v_{k}}{\sigma} < 0.78$. In all other cases, the decay of the BH's orbit is much slower than predicted.

The discrepancy between classical DF theory and the orbital decay of BHs in our simulations can be attributed the initial assumptions made by \citet{chandrasekhar43} when developing a formalism for dynamical friction and the additional assumptions that need to be made in order to generate an analytical prediction for an orbital decay rate \citep[e.g.][]{chatterjee02, gualandris08}. More specifically, DF theory does not account for inhomogeneous background densities, $v_{k} > \sigma(a)$, $r > a$, and two-body interactions. The latter of which can even result in BHs gaining energy which will further delay orbital decay.

To account for these additional factors, we first attempted to determine a correction factor that can be applied to the exponential decay rate that one would calculate assuming the BH is orbiting within a homogeneous stellar field, $v_{k} > \sigma(a)$ and $r > a$. However we find that the correction factor, which is independent of BH mass, can only reproduce the orbital decay of a BH for select cases that have $\frac{v_{k}}{\sigma} \le 1.6$. Furthermore, the correction factor has a very weak dependence on $\frac{v_{k}}{\sigma}$. For all of our $\frac{v_{k}}{\sigma} \le 1.6$ models, we find that a linear decay rate can also accurately model a BHs decay and in some cases is even preferred. For $\frac{v_{k}}{\sigma} > 1.6$, we find that the BH orbits no longer reflect a damped harmonic oscillator and can only be modelled assuming a linear decay rate. Furthermore, the linear decay rate $c_L$ scales linearly with both $m_{bh}$ and $\frac{v_{k}}{\sigma}$. To directly apply our linear formalism to GCs and NSCs, the influence of stars over an entire mass spectrum, the presence of a central SMBH, and a wider range of cluster masses and density profiles must also be taken into account and requires further study in future work. 

Knowing the actual decay time of kicked black holes, which we have shown to be significantly longer than DF theory would predict, directly affects the ability of stellar, intermediate and supermassive BHs to form and grow in dense stellar environments through gravitational wave producing mergers. In fact, the region of the $m_{bh}$ - $\frac{v_{k}}{\sigma}$ parameter space explored in this study, which is of particular interest for the BHs that have been found to emit gravitational waves upon formation, is in need of a new and dynamically motivated approach for modelling the long term evolution of BHs in star cluster environments. Hence future studies will attempt to isolate the effects of inhomogeneity and the number of two-body interactions on our model.

\section*{Acknowledgments}

This work was made possible by the facilities of the Shared Hierarchical Academic Research Computing Network (SHARCNET:www.sharcnet.ca) and Compute/Calcul Canada. N.~W.~C.~L. gratefully acknowledges support from the American Museum of Natural History and the Richard Guilder Graduate School, specifically the Kalbfleisch Fellowship Program, as well as support from a National Science Foundation Award No. AST 11-09395. The authors would also like to kindly thank Mordecai-Mark Mac Low and Enrico Vesperini for useful discussions and suggestions. 


\bsp

\label{lastpage}

\end{document}